\documentclass[preprint,showpacs,preprintnumbers,showkeys,amsmath,amssymb]{revtex4}

\usepackage{graphicx}
\usepackage{dcolumn}
\usepackage{bm}

\begin{document}


\title{Brewster angle for anisotropic materials from the extinction theorem}

\author{Weixing Shu}\thanks{Corresponding author. $E$-$mail$ $address$: wxshuz@gmail.com.}
\author{Zhongzhou Ren}
\author{Hailu Luo}
\author{Fei Li}

\affiliation{ Department of Physics, Nanjing University, Nanjing
210008, China}

\begin{abstract}
We explore the physical origin of Brewster angle in the external
and internal reflections associated with an anisotropic material.
We obtain the expressions of the reflected fields and the
existence condition of Brewster angle by using the extinction
theorem. It is found that the Brewster angle will occur if the
total contribution of the anisotropic material's electric and
magnetic dipoles to the reflection field becomes zero. In internal
reflection, the requirements on the material parameters
${\boldsymbol \varepsilon}$ and ${\boldsymbol \mu}$ for Brewster
angle are the same as those in external reflection, and the
Brewster angle is just the refraction angle in external reflection
at the incidence of external Brewster angle. In contrast to the
conventional isotropic medium, an anisotropic material can exhibit
Brewster angle for both TE and TM waves due to its anisotropy. The
results of the present paper are applicable to anisotropic
dielectric and magnetic materials, including metamaterials.
\end{abstract}

\maketitle

\section{Introduction}\label{Introduction}
As well known, Brewster angle is the angle of incidence at which
the incident light is reflected without the polarization component
parallel to the plane of incidence. At the Brewster angle, the
reflected light is perpendicular to the refracted light. The
physics accounting for such a phenomenon is that the vibration of
electrons in the second medium can not generate the reflected
light which travels perpendicular to the transmitted light
\cite{Born}. These conclusions only hold for isotropic dielectric
materials. Since conventional transparent isotropic materials can
be regarded nonmagnetic, these conclusions are applicable to them.

However, the advent of a new kind of artificial materials, named
as left-handed materials, changes the situation. The left-handed
material was hypothesized by Veselago \cite{Veselago1968} and can
exhibit many exotic electromagnetic properties, among which the
most well known is the negative refraction. Since the negative
refraction was experimentally observed in a structured
metamaterial composed of arrays of conducting split ring
resonators (SRRs) and wires \cite{Shelby2001}, the left-handed
material has sparked great interest
\cite{Pendry2000,Lindell2001,Smith2003,Smith2004,Hu2002,Luo2002,Belov2003,Lakhtakia2004,Lakhtakia2006,Luo2005,Luo2006}.
Metamaterials have been explored to exhibit Brewster angle not
only for TM (traverse magnetic) waves, but also for TE (traverse
electric) waves \cite{Zhou2003,Grzegorczyk2005}. Then, one
enquires naturally: How on earth do TE waves exhibit Brewster
angle in metamaterials? Whether the mechanism of Brewster angle
for TE waves is the same as that for TM waves, i.e., just
described above?

It is well accepted that the molecular optics theory can give much
deeper physical insight into the interaction of electromagnetic
wave with material than do Maxwell theory
\cite{Feynman1963,Wolf1972,Reali1982,Karam1997,Doyle1985}. But
such an approach is less frequently employed because it involves
integral-differential equations difficult to solve. Recently, Lai
\textit{et al.} \cite{Lai2002} used the method of superposition of
retarded field to discuss the reflection and refraction law of
electromagnetic wave incident on an isotropic medium. Along that
way, Fu \textit{et al.} \cite{Fu2005} explored the Brewster
condition for light incident from the vacuum onto an isotropic
material with negative index. Since the metamaterials are actually
anisotropic, it is necessary to generalize the Brewster condition
from the isotropic material to the anisotropic material. At the
same time, in most work dealing with the interaction of
electromagnetic wave with material by the molecular optics theory,
often considered is the wave incident from the vacuum into a
dielectric material \cite{Feynman1963,Reali1982,Karam1996}, but
the case of wave incident from the vacuum on a magnetic material,
or from a material into vacuum is rarely investigated.

The purpose of this paper is to present a detailed investigation
on the mechanism of Brewster angle in external and internal
reflections associated with an anisotropic dielectric-magnetic
material. We use the extinction theorem of the molecular optics to
derive the reflected fields and the existence condition of
Brewster angle. We find that Brewster angle will occur if the
contributions of the electric and magnetic dipoles to the
reflected field add up to zero. We also study in detail the
impacts of  the material parameters ${\boldsymbol \varepsilon}$
and ${\boldsymbol \mu}$ on the Brewster angle. The results extend
the conclusions about Brewster angle in isotropic materials
\cite{Fu2005,Lai2002,Reali1982,Doyle1985} and can provide
references in manufacturing materials for specific purposes, such
as making polarization devices. The conclusions also provide a new
and deep look on those obtained by Maxwell theory \cite{Shen2006}.

\section{Brewster angle in external reflection}\label{}
In molecular optics theory, a bulk material can be regarded as a
collection of molecules (or atoms) embedded in the vacuum. Under
the action of an incident field, the molecules oscillate as
electric and magnetic dipoles and emit radiations. The radiation
field and the incident field interact to form the new transmitted
field in the material and the reflection field outside the
material~\cite{Born}.

In this section, we first employ the Ewald-Oseen extinction
theorem to deduce the expressions of radiated fields generated by
dipoles in the external reflection of waves incident on an
anisotropic dielectric-magnetic material. Then we study the
Brewster angle condition and discuss the results.

\subsection{Extinction theorem and external reflection}\label{}
\begin{figure}[t]
\centering
\includegraphics[width=8cm]{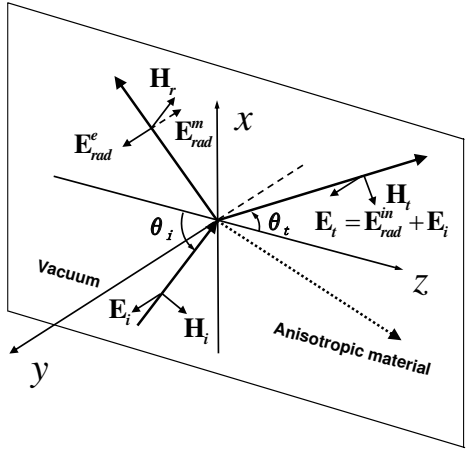}
\caption{\label{shuFig2}Schematic diagram for how the reflected
and transmitted fields of TE waves are generated by the incident
field and radiated fields of dipoles. In the vacuum, the reflected
field ${\bf E}_r={\bf E}^e_{r0}+{\bf E}^m_{r0}$, while the
transmitted field ${\bf E}_t={\bf E}_i+{\bf E}^{in}_{rad}$ in the
material. When ${\bf E}^e_{r0}+{\bf E}^m_{r0}=0$, ${\bf E}_{r0}=0$
and Brewster angle occurs for TE waves. The dotted line is the
transmitted wave in negative refraction.}
\end{figure}

Let a monochromatic electromagnetic field of ${\bf E}_i={\bf
E}_{i0} \exp{(i{\bf k}_i\cdot {\bf r}-i\omega t)}$ and ${\bf
H}_i={\bf H}_{i0} \exp{(i{\bf k}_i\cdot {\bf r}-i\omega t)}$
incident from the vacuum on an anisotropic material filling the
semi-infinite space $z>0$ with ${\bf k}_i=k_{ix}\hat{{\bf
x}}-k_{iz}\hat{{\bf z}}$. The $x-z$ plane is the plane of
incidence and the schematic diagram is shown in
Fig.~\ref{shuFig2}. Since the material responds linearly,  all the
fields have the same dependence of $\exp{(-i\omega t)}$ which will
be omitted subsequently for simplicity. We assume that the
reflected fields are ${\bf E}_r={\bf E}_{r0} \exp{(i{\bf k}_r\cdot
{\bf r})}$ and ${\bf H}_r={\bf H}_{r0} \exp{(i{\bf k}_r\cdot {\bf
r})}$, and the transmitted fields are ${\bf E}_t={\bf E}_{t0}
\exp{(i{\bf k}_t\cdot {\bf r})}$ and  ${\bf H}_t={\bf H}_{t0}
\exp{(i{\bf k}_t\cdot {\bf r})}$, where ${\bf k}_r=k_{rx}\hat{{\bf
x}}-k_{rz}\hat{{\bf z}}$ and ${\bf k}_t=k_{tx}\hat{{\bf
x}}+k_{tz}\hat{{\bf z}}$. The permittivity and permeability
tensors of the anisotropic material are simultaneously diagonal in
the principal coordinate system,
$\boldsymbol{\varepsilon}=\hbox{diag}[\varepsilon_x,
\varepsilon_y, \varepsilon_z], ~~\boldsymbol{\mu}=\hbox{diag}
[\mu_x, \mu_y, \mu_z]$.

Inside the anisotropic material, the incident field drive the
dipoles to oscillate and radiate. The electric fields  radiated by
electric dipoles and magnetic dipoles are respectively decided by
\cite{Born}
\begin{eqnarray}\label{Ee_dipole_field}
&&{\bf
E}^{e}_{rad}=\nabla(\nabla\cdot{\bf{\Pi}}_e)-\varepsilon_0\mu_0\frac{\partial^2{\bf{\Pi}}_e}{\partial
t^2},\\
&&{\bf
E}^{m}_{rad}=-\mu_0\nabla\times\frac{\partial{\bf{\Pi}}_m}{\partial
t}.\label{Em_dipole_field}
\end{eqnarray}
And the  magnetic fields generated by magnetic dipoles and
electric dipoles  are
\begin{eqnarray}\label{Hm_dipole_field}
&&{\bf
H}^{m}_{rad}=\nabla(\nabla\cdot{\bf{\Pi}}_m)-\varepsilon_0\mu_0\frac{\partial^2{\bf{\Pi}}_m}{\partial
t^2},\\
&&{\bf
H}^{e}_{rad}=\varepsilon_0\nabla\times\frac{\partial{\bf{\Pi}}_e}{\partial
t}.\label{He_dipole_field}
\end{eqnarray}
Here ${\bf \Pi}_e$ and ${\bf \Pi}_m$ are the Hertz vectors,
\begin{eqnarray}\label{Pi_e}
&&{\bf \Pi}_e({\bf r})=\int \frac{{\bf P}({\bf
r'})}{\varepsilon_0}G({\bf r}-{\bf r}')\hbox{d}{\bf r}',\\
&&{\bf \Pi}_m({\bf r})=\int {\bf M}({\bf r'})G({\bf r}-{\bf
r}')\hbox{d}{\bf r}'.\label{Pi_m}
\end{eqnarray}
${\bf P}$ is the dipole moment density of electric dipoles and
${\bf M}$ is that of magnetic dipoles, which are related to the
transmitted fields by ${\bf
P}=\varepsilon_0\boldsymbol{\chi}_e\cdot{\bf E}_t$ and  ${\bf
M}=\boldsymbol{\chi}_m\cdot{\bf H}_t$, where the electric
susceptibility
$\boldsymbol{\chi}_e=(\boldsymbol{\varepsilon}/{\varepsilon_0})-1$
and the magnetic susceptibility
$\boldsymbol{\chi}_m=(\boldsymbol{\mu}/{\mu_0})-1$. The Green
function is $G({\bf r}-{\bf r}')=\exp{(ik_i|{\bf r}-{\bf
r}'|)}/(4\pi|{\bf r}-{\bf r}'|)$. To evaluate the Hertz vectors,
we firstly represent the Green function in the Fourier form. Then,
inserting it into Eqs.~(\ref{Pi_e}) and (\ref{Pi_m}) and using the
delta function definition and contour integration method
\cite{Reali1982}, the Hertz vectors can be evaluated as
\begin{equation}\label{Pi_e_ex}
{\bf \Pi}_e=\left\{
\begin{array}{cc}
\displaystyle  -\frac{\boldsymbol{\chi}_e\cdot{\bf
E}_{t0}\exp{(ik_{tx}x-ik_{1z}z})}{2k_{1z}(k_{1z}+k_{tz})},&-\infty<z<0\\
\displaystyle  \frac{\boldsymbol{\chi}_e\cdot{\bf
E}_{t0}\exp{(i{\bf k}_{1}\cdot{\bf
r})}}{2k_{1z}(k_{1z}-k_{tz})}+\frac{\boldsymbol{\chi}_e\cdot{\bf
E}_{t0}\exp{(i{\bf k}_t\cdot{\bf r})}}{k_{t}^2-k_{i}^2},&0\leq
z<\infty
\end{array}\right.
\end{equation}
\begin{equation}\label{Pi_m_ex}
{\bf \Pi}_m=\left\{
\begin{array}{cc}
\displaystyle  -\frac{\boldsymbol{\chi}_m\cdot{\bf
H}_{t0}\exp{(ik_{tx}x-ik_{1z}z)}}
{2k_{1z}(k_{1z}+k_{tz})},&-\infty<z<0\\
\displaystyle  \frac{\boldsymbol{\chi}_m\cdot{\bf
H}_{t0}\exp{(i{\bf k}_{1}\cdot{\bf
r})}}{2k_{1z}(k_{1z}-k_{tz})}+\frac{\boldsymbol{\chi}_m\cdot{\bf
H}_{t0}\exp{(i{\bf k}_{t}\cdot{\bf r})}}{k_{t}^2-k_{i}^2},&0\leq
z<\infty
\end{array}\right.
\end{equation}
where ${\bf k}_1=k_{tx}\hat{{\bf x}}+k_{1z}\hat{{\bf z}}$,
$k_{1z}^2=k_i^2-k_{tx}^2$, and we have used the Faraday's law
${\bf H}_t=({\bf k}_t\times{\bf E}_t)/(\omega {\boldsymbol \mu})$
which  can also be established by the molecular theory.

Following the extinction theorem, the incident field is
extinguished inside the material and is replaced by the
transmitted field \cite{Born}. Then, we get
\begin{equation}\label{E_t}
{\bf E}_t={\bf E}^{e}_{rad}+{\bf E}^m_{rad}+{\bf E}_i.
\end{equation}
Using Eqs.~(\ref{Pi_e_ex}) and (\ref{Pi_m_ex}) and inserting
Eqs.~(\ref{Ee_dipole_field}) and (\ref{Em_dipole_field}) into
Eq.~(\ref{E_t}), we come to the following conclusions.

(1). Comparing terms of the phase factor $\exp{(i{\bf
k}_{1}\cdot{\bf r})}$ in Eq.~(\ref{E_t}), we know that ${\bf
k}_1={\bf k}_i$ and $k_{ix}=k_{tx}$. This is just the Snell's law:
$k_{i}\sin\theta_i=k_{t}\sin\theta_t$.

(2). At the same time, the incident field can be written in terms
of the transmitted field
\begin{equation}\label{E_i=}
{\bf E}_{i0}=\frac{{\bf k}_i\times[{\bf k}_i\times
(\boldsymbol{\chi}_e\cdot{\bf
E}_{t0})]}{2k_{iz}(k_{iz}-k_{tz})}+\frac{{\bf
k}_i\times\{\boldsymbol{\chi}_m\cdot[{\mu_0}{\boldsymbol{\mu}}^{-1}\cdot({\bf
k}_t\times {\bf E}_{t0})]\}}{2k_{iz}(k_{iz}-k_{tz})}.
\end{equation}
Equation~(\ref{E_i=}) is actually the expression of the
Ewald-Odseen extinction theorem. It shows quantitatively  how the
radiation field of dipoles extinguish the incident field.

(3). The terms with the phase factor $\exp{(i{\bf k}_{t}\cdot{\bf
r})}$ in Eq.~(\ref{E_t}) yields the dispersion relation
\begin{equation}
\frac{k_{tx}^2}{\mu_z \varepsilon_y}+\frac{k_{tz}^2}{\mu_x
\varepsilon_y }=\omega^2 \label{D1}
\end{equation}
for TE waves. In order to guarantee $k_{tz}$ real, it requires
that $\varepsilon_{y}{\mu_z}<\varepsilon_{0}{\mu_0}\cap
{\mu_x}{\mu_z}<0$, or $\varepsilon_{y}{\mu_z}>0\cap
{\mu_x}{\mu_z}>0$. In addition, there will exist a critical angle
of incidence $\theta^{TE}_C= \sin
^{-1}\sqrt{{\mu_z}\varepsilon_{y}/{\varepsilon_0}\mu_0}$ if
$0<{\mu_z}\varepsilon_{y}<{\varepsilon_0}\mu_0$. Outside the
anisotropic material, the contributions from the electric and
magnetic dipoles form the reflected field ${\bf E}_{r}$. Applying
Eqs.~(\ref{Pi_e_ex}) and (\ref{Pi_m_ex}) for $z<0$ to
Eqs.~(\ref{Ee_dipole_field}) and (\ref{Em_dipole_field}), we
obtain
\begin{eqnarray}\label{E_r=}
{\bf E}_{r0}&=&{\bf E}^e_{r0}+{\bf E}^m_{r0}\nonumber\\
&=&\frac{{\bf k}_r\times[{\bf k}_r\times
(\boldsymbol{\chi}_e\cdot{\bf
E}_{t0})]}{2k_{iz}(k_{iz}+k_{tz})}+\frac{{\bf
k}_r\times\{\boldsymbol{\chi}_m\cdot[{\mu_0}{\boldsymbol{\mu}}^{-1}\cdot({\bf
k}_t\times {\bf E}_{t0})]\}}{2k_{iz}(k_{iz}+k_{tz})}.
\end{eqnarray}
where ${\bf k}_r=k_{ix}\hat{{\bf x}}-k_{iz}\hat{{\bf z}}$.
Equations~(\ref{E_i=}) and (\ref{E_r=}) hold for both TE and TM
waves. And we obtain the reflection coefficient $R_E(={ E}_{r0}/{
E}_{i0})$ and the transmission coefficient $T_E(={E}_{t0}/{
E}_{i0})$ for TE waves
\begin{equation}\label{RE}
R_E=\frac{\mu_xk_{iz}-\mu_0k_{tz}}{\mu_xk_{iz}+\mu_0k_{tz}}
,~~~T_E=\frac{2\mu_xk_{iz}}{\mu_xk_{iz}+\mu_0k_{tz}}.
\end{equation}

Analogously, we can derive the incident magnetic field
\begin{equation}\label{H_i=}
{\bf H}_{i0}=\frac{{\bf k}_i\times[{\bf k}_i\times
(\boldsymbol{\chi}_m\cdot{\bf
H}_{t0})]}{2k_{iz}(k_{iz}-k_{tz})}+\frac{{\bf
k}_i\times\{\boldsymbol{\chi}_e\cdot[{\varepsilon_0}{\boldsymbol{\varepsilon}}^{-1}\cdot({\bf
k}_t\times {\bf H}_{t0})]\}}{2k_{iz}(k_{iz}-k_{tz})},
\end{equation}
the reflected magnetic field
\begin{eqnarray}\label{H_r=}
{\bf H}_{r0}&=&{\bf H}^m_{r0}+{\bf H}^e_{r0}\nonumber\\
&=&\frac{{\bf k}_r\times[{\bf k}_r\times
(\boldsymbol{\chi}_m\cdot{\bf
H}_{t0})]}{2k_{iz}(k_{iz}+k_{tz})}+\frac{{\bf
k}_r\times\{\boldsymbol{\chi}_e\cdot[{\varepsilon_0}{\boldsymbol{\varepsilon}}^{-1}\cdot({\bf
k}_t\times {\bf H}_{t0})]\}}{2k_{iz}(k_{iz}+k_{tz})},
\end{eqnarray}
and the dispersion relation
\begin{equation}
\frac{k_{tx}^2}{\varepsilon_z \mu_y}+\frac{k_{tz}^2}{\varepsilon_x
\mu_y }=\omega^2 \label{D2}
\end{equation}
for TM waves. To ensure $k_{tz}$ real, it needs that
$\varepsilon_{z}{\mu_y}<\varepsilon_{0}{\mu_0}\cap
{\varepsilon_x}{\varepsilon_z}<0$, or
$\varepsilon_{z}{\mu_y}>0\cap {\varepsilon_x}{\varepsilon_z}>0$.
In addition, there will be a critical angle of incidence
$\theta^{TM}_C= \sin
^{-1}\sqrt{{\mu_y}\varepsilon_{z}/{\varepsilon_0}\mu_0}$ if
$0<{\mu_y}\varepsilon_{z}<{\varepsilon_0}\mu_0$. And we obtain the
reflection coefficient $R_H(={ H}_{r0}/{ H}_{i0})$ and the
transmission coefficient $T_H(={ H}_{t0}/{ H}_{i0})$ for TM waves
as
\begin{equation}\label{RH}
R_H=\frac{\varepsilon_xk_{iz}-\varepsilon_0k_{tz}}{\varepsilon_xk_{iz}+\varepsilon_0k_{tz}},
~~~T_H=\frac{2\varepsilon_xk_{iz}}{\varepsilon_xk_{iz}+\varepsilon_0k_{tz}},
\end{equation}
respectively.

Obviously, Eqs.~(\ref{RE}) and (\ref{RH}) are in agreement with
the results obtained by the formal approach of Maxwell's
equations. At the same time, we can see that the extinction
theorem plays the role of the boundary conditions in Maxwell
approach.

\subsection{The origin of Brewster angle and the impact of $\boldsymbol{\varepsilon}$ and $\boldsymbol{\mu}$}
Let us now apply the results just obtained to study the origin of
Brewster angle in the reflection of waves incident on the
anisotropic material.

If the power reflectivity $r=|R|^2=0$, there is no reflected wave
and the incident angle is named as Brewster angle \cite{Kong2000}.
Now we discuss TE and TM waves separately. In order to meet
$r_E=|{\bf E}_{r0}/{\bf E}_{i0}|^2=0$, it requires that ${\bf
E}_{r0}=0$ in Eq.~(\ref{E_r=}), i.e.,
\begin{equation}\label{E_r=0_E1}
\frac{(1-{\varepsilon_y}/{\varepsilon_0})k_i^2-k_{ix}k_{tx}(1-{\mu_0}/{\mu_z})
+k_{iz}k_{tz}(1-{\mu_0}/{\mu_x})}{2k_{iz}(k_{iz}+k_{tz})}=0.
\end{equation}
From Eq.~(\ref{E_r=0_E1}), it follows that if
\begin{equation}\label{Brewster_E_condition}
0<\frac{1-\frac{\mu_0\varepsilon_y}{\mu_x\varepsilon_0}}{1-\frac{\mu_0\mu_0}{\mu_x\mu_z}}<1
\end{equation}
the Brewster angle for TE waves is
\begin{equation}\label{Brewster_E}
\theta_{B}^{TE}=\sin ^{-1}
\sqrt{\frac{1-\frac{\mu_0\varepsilon_y}{\mu_x\varepsilon_0}}{1-\frac{\mu_0\mu_0}{\mu_x\mu_z}}}.
\end{equation}
To realize $r_H=|{\bf H}_{r0}/{\bf H}_{i0}|^2=0$, it is needed
that ${\bf H}_{r0}=0$, that is,
\begin{equation}\label{E_r=0_H1}
\frac{(1-{\mu_y}/{\mu_0})k_i^2-k_{ix}k_{tx}(1-{\varepsilon_0}
/{\varepsilon_z})+k_{iz}k_{tz}(1-{\varepsilon_0}/{\varepsilon_x})}{2k_{iz}(k_{iz}+k_{tz})}=0.
\end{equation}
Similarly, we conclude that under the condition
\begin{equation}\label{Brewster_H_condition}
0<\frac{1-\frac{\varepsilon_0\mu_y}{\varepsilon_x\mu_0}}{1-\frac{\varepsilon_0\varepsilon_0}{
\varepsilon_x\varepsilon_z}}<1,
\end{equation}
there exists a Brewster angle for TM waves
\begin{equation}\label{Brewster_H}
\theta_{B}^{TM}=\sin ^{-1}
\sqrt{\frac{1-\frac{\varepsilon_0\mu_y}{\varepsilon_x\mu_0}}{1-\frac{\varepsilon_0\varepsilon_0}{
\varepsilon_x\varepsilon_z}}}.
\end{equation}

We find that the Brewster condition of TE waves is only related to
three components of the material parameters ${\boldsymbol
\varepsilon}$ and ${\boldsymbol \mu}$, while that of TM waves
depends on the other three components. Therefore, we can let the
anisotropic material exhibit Brewster angles for TE, or TM, or
both waves through choosing appropriate ${\boldsymbol
\varepsilon}$ and ${\boldsymbol \mu}$. This is in sharp contrast
with the regular isotropic material case where only one of TE and
TM waves can exhibit Brewster angle.

\begin{figure}
\centering
\includegraphics[width=8cm]{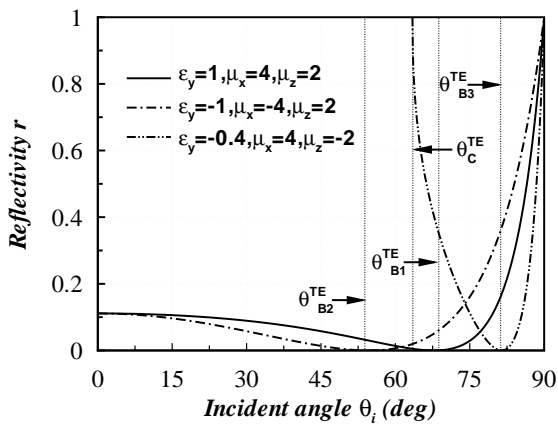}%
\caption{\label{r}Reflectivity $r$ of TE waves as a function of
the incidence angle $\theta_i$ at the interface between vacuum and
an \textit{cutoff}, \textit{never cutoff}, and
\textit{anti-cutoff} anisotropic materials. There exists a
Brewster angle for the three cases. Note that there exists a
critical angle of incidence $\theta^{TE}_C$ in reflection on
\textit{anti-cutoff} anisotropic material.}
\end{figure}

\begin{figure}
\centering
\includegraphics[width=8cm]{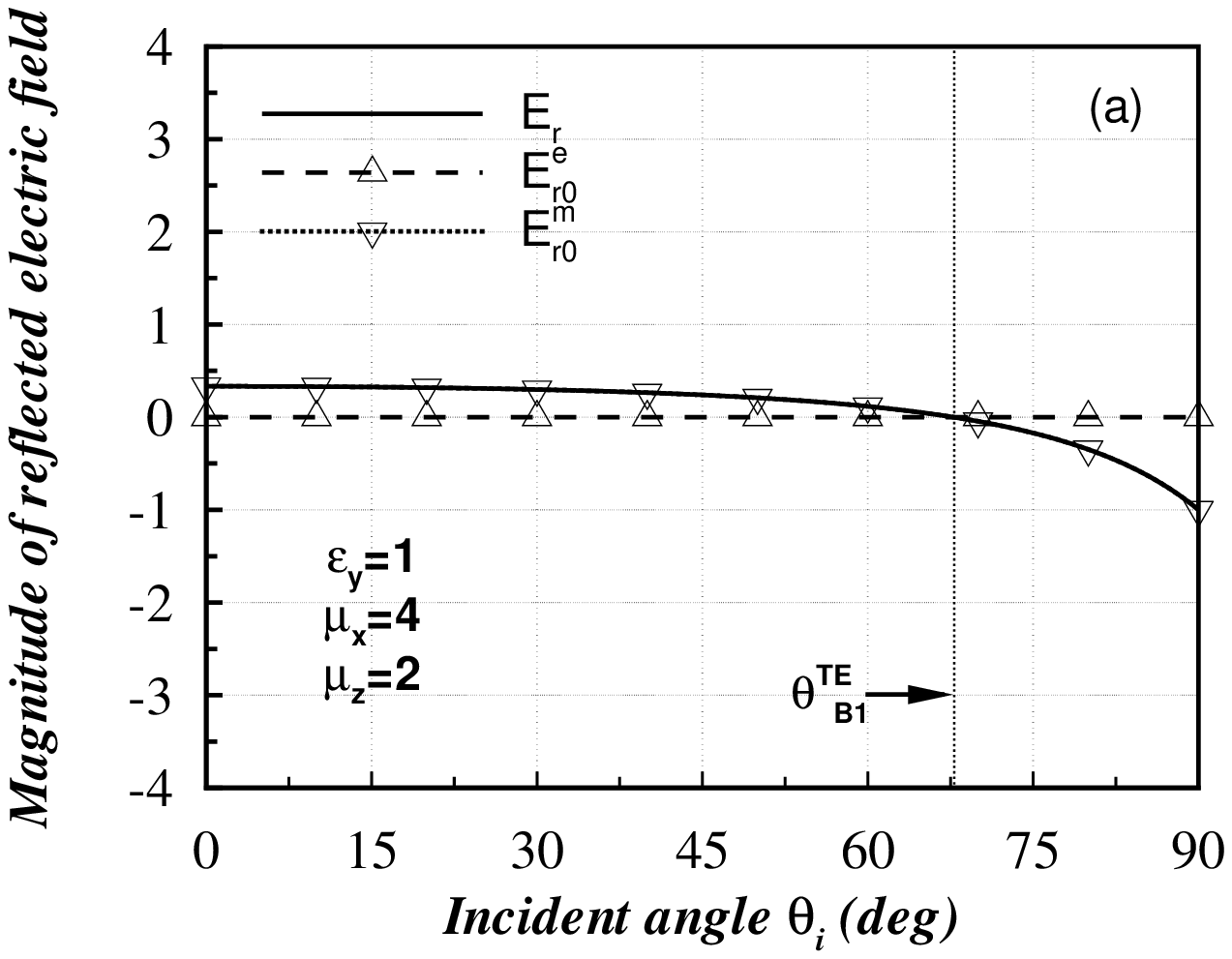}
\includegraphics[width=8cm]{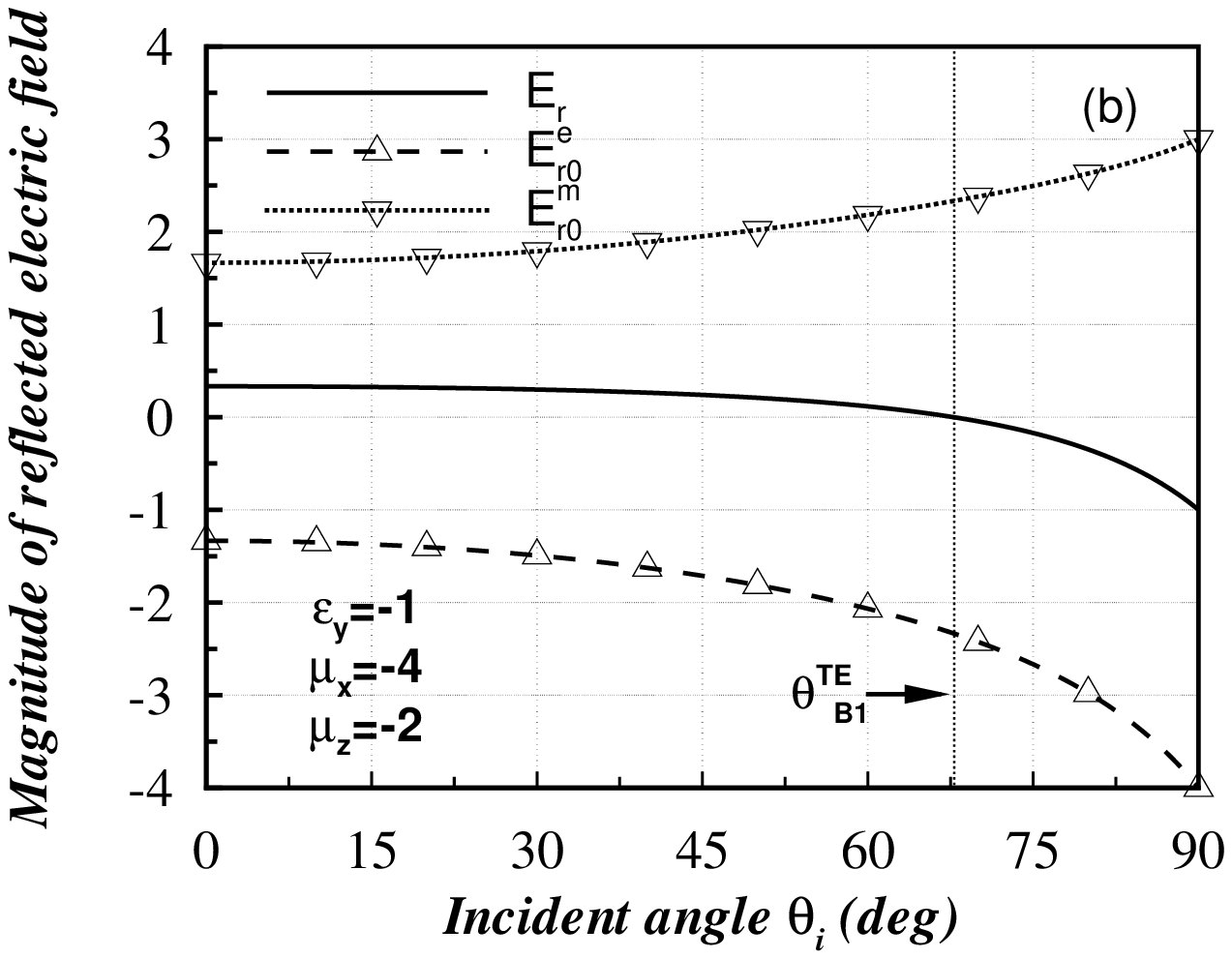}
\caption{\label{cutoff}Reflected field magnitudes, normalized by
the corresponding incident field magnitudes, for TE wave incident
from vacuum into a \textit{cutoff} anisotropic material. Since
$\varepsilon_y=\varepsilon_0$, the radiated electric fields of
electric dipoles ${\bf E}^e_{r0}=0$, then ${\bf E}_{r0}={\bf
E}^m_{r0}$. When ${\bf E}^m_{r0}=0$ Brewster angle $\theta_B^{TE}$
appears, as shown in (a).}
\end{figure}

\begin{figure}
\centering
\includegraphics[width=8cm]{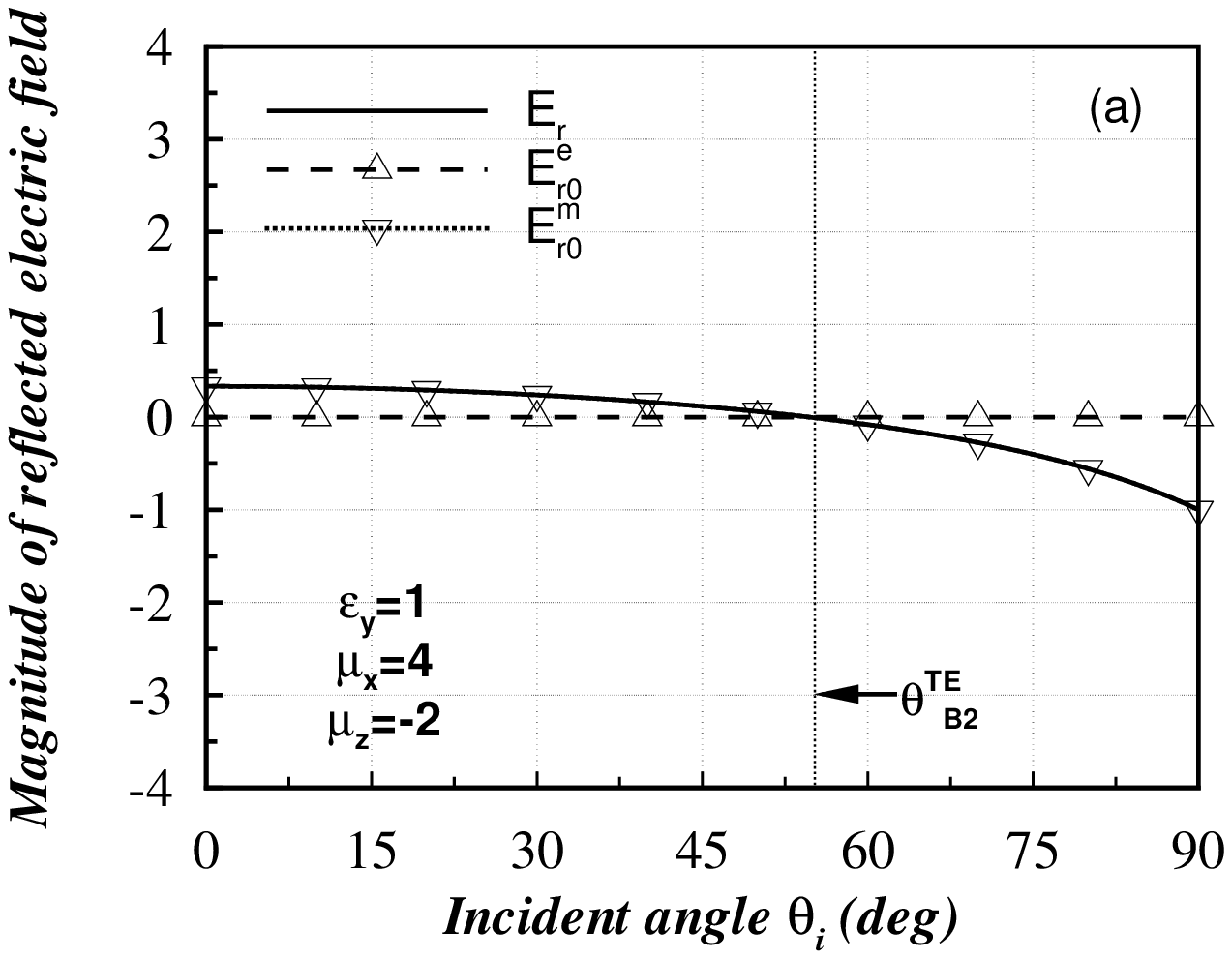}
\includegraphics[width=8cm]{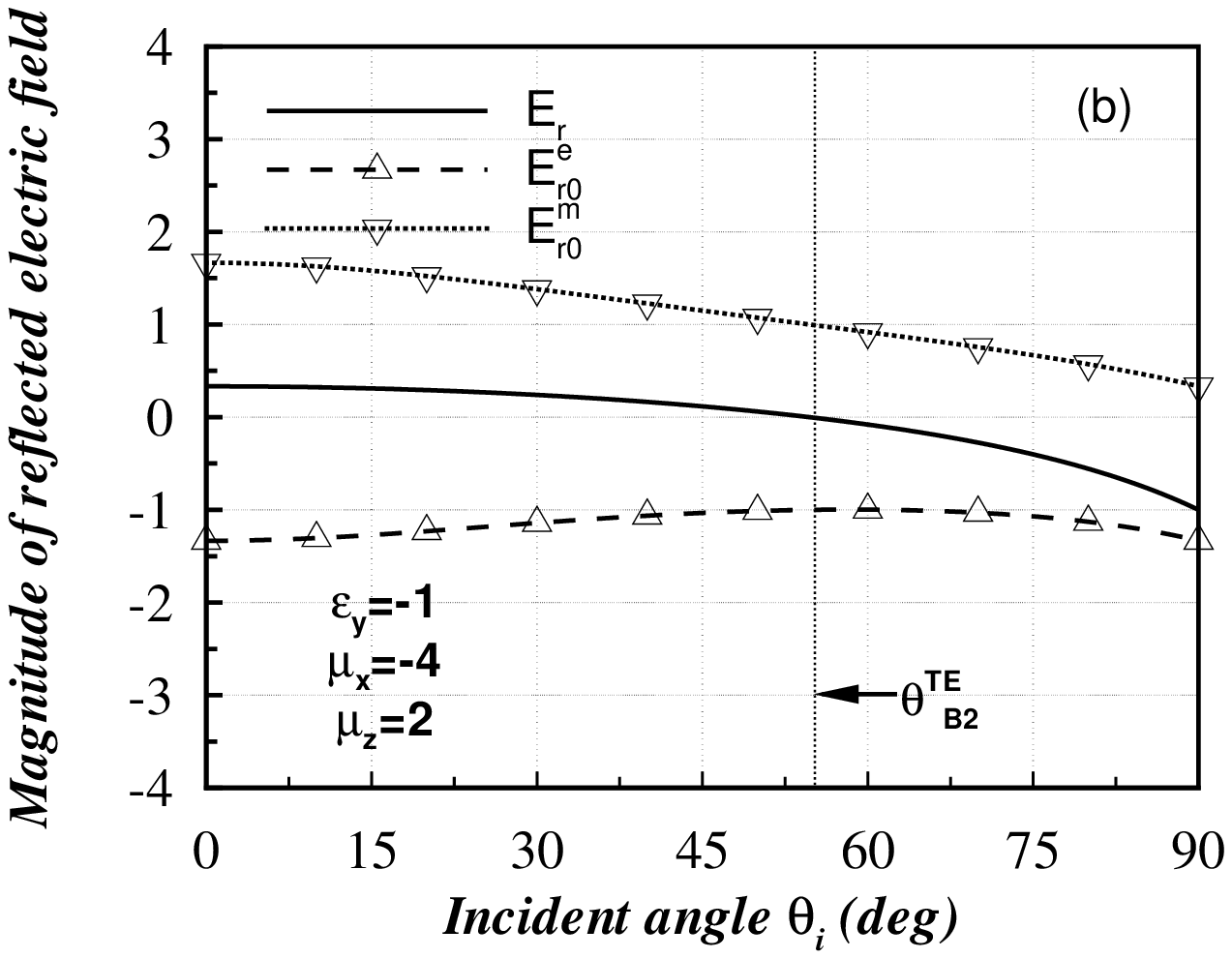}
\caption{\label{never cutoff}Reflected field magnitudes,
normalized by the corresponding incident field magnitudes, for TE
wave incident from vacuum into a \textit{never cutoff} anisotropic
material.}
\end{figure}

\begin{figure}
\centering
\includegraphics[width=8cm]{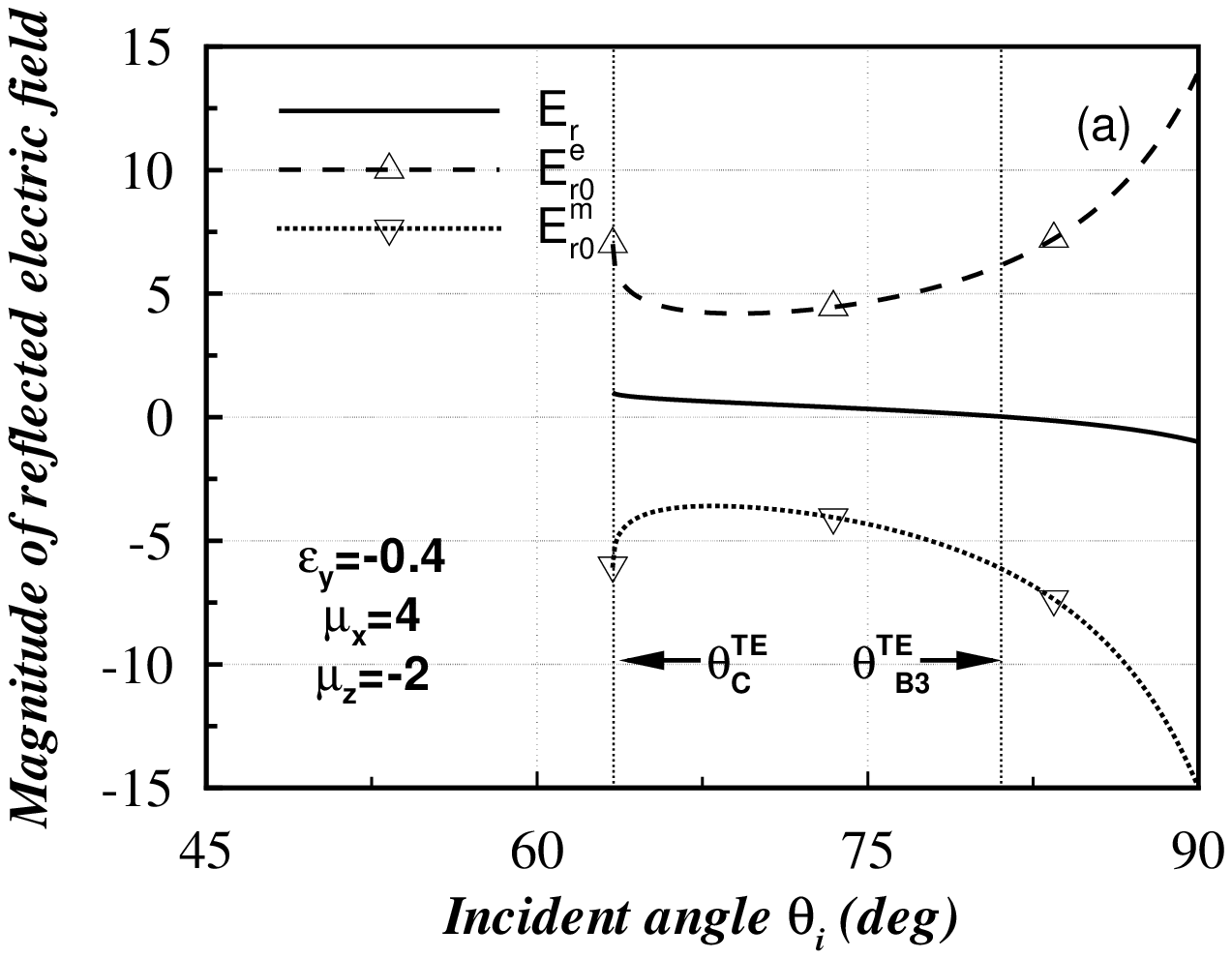}
\includegraphics[width=8cm]{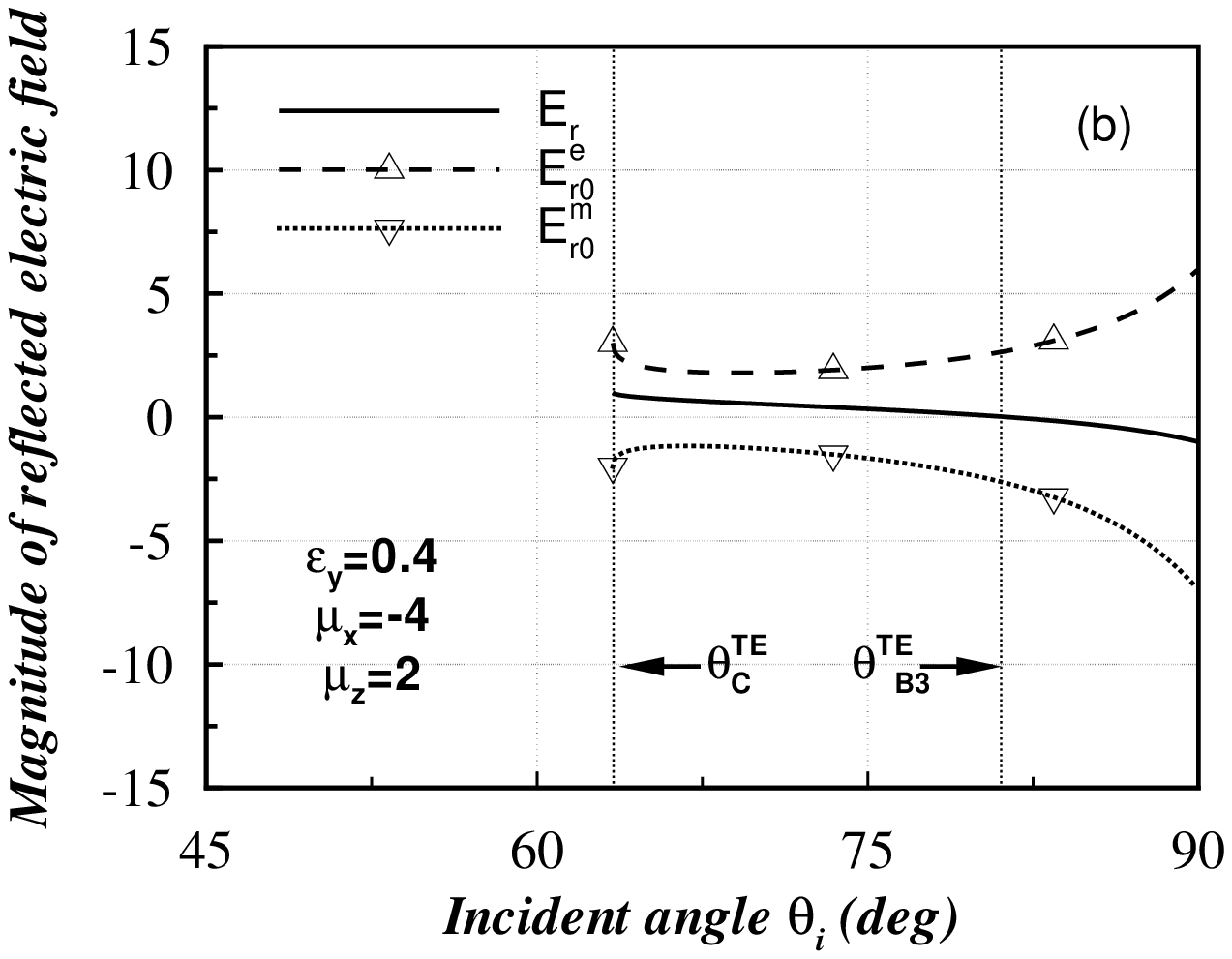}
\caption{\label{anti-cutoff}Reflected field magnitudes, normalized
by the corresponding incident field magnitudes, for TE wave
incident from vacuum into an \textit{anti-cutoff} anisotropic
material. There exist a critical angle $\theta^{TE}_C$ for
incidence and $\theta^{TE}_C>\theta^{TE}_{B3}$, which is called as
the inversion of critical angle.}
\end{figure}

There are different sign combination of ${\boldsymbol
\varepsilon}$ and ${\boldsymbol \mu}$ for the anisotropic
materials. According to the form of dispersion relation Smith
\textit{et al.} classify the anisotropic material into three types
: \textit{cutoff}, \textit{never cutoff}, and \textit{anti-cutoff}
\cite{Smith2003}. We give an example of the reflectivity for wave
incident into each type in Fig.~\ref{r}.
Clearly, we see that there is Brewster angle in the reflection of
TE (the case of TM waves can be discussed similarly). To explain
the Brewster condition vividly, we illustrate in
Figs.~\ref{cutoff}, \ref{never cutoff} and \ref{anti-cutoff} the
magnitudes of radiation fields for the examples in Fig.~\ref{r}.
It is clear that when the total radiated field of electric and
magnetic dipoles is zero, i.e., ${\bf E}_{r0}^e+{\bf E}_{r0}^m=0$,
the Brewster angle occurs. Each of the three classes of media has
two subtypes: one positive (fig(a))and one negative (fig(b))
refracting. Comparing the two subtypes of each figure, one can
find that the reflectivity is the same, but the field magnitudes
${\bf E}^e_{r0}$ and ${\bf E}^m_{r0}$ are totally different
because signs of ${\boldsymbol \varepsilon}$ and ${\boldsymbol
\mu}$ are reversed. Even if one element's sign changes, such as
$\mu_z$ in Figs.~\ref{cutoff}(b) and \ref{never cutoff}(b), ${\bf
E}^e_{r0}$ and ${\bf E}^m_{r0}$ alter accordingly. If one
element's magnitude and sign change, then not only the  magnitude
but also the phase of the radiated field can change, such as in
Figs.~\ref{never cutoff}(b) and \ref{anti-cutoff}(b).

In the next step, we study the impacts of ${\boldsymbol
\varepsilon}$ and ${\boldsymbol \mu}$ on Brewster angles for TE
and TM waves.

(1).~TE waves. In Eq.~(\ref{E_r=0_E1}) the first term denotes the
contribution of electric dipoles ${\bf E}_{r0}^e$, and the other
two stand for the contributions of magnetic dipoles ${\bf
E}_{r0}^{m}$. Obviously, the condition
(\ref{Brewster_E_condition}) is only connected with
${\varepsilon_y}/{\varepsilon_0}$, ${\mu_0}/{\mu_z}$ and
${\mu_0}/{\mu_x}$, which determine the magnitudes of contributions
of the dipoles, i.e., ${\bf E}_{r0}^e$ and ${\bf E}_{r0}^m$. The
relevant points to note are as follows. (i) If
$\varepsilon_y=\varepsilon_0$, then ${\bf E}_{r0}^{e}=0$. From the
condition Eq.~(\ref{E_r=0_E1}) we know the angle between the
reflection and refraction waves satisfies
\begin{equation}\label{}
\cos\theta=\frac{k_{ix}k_{tx}}{k_ik_t}\frac{\mu_0(\mu_x-\mu_z)}{\mu_z(\mu_x-\mu_0)}.
\end{equation}
Obviously, if $\mu_x\neq\mu_z$, then $\cos\theta\neq0$ and ${\bf
k}_r$ is not perpendicular to ${\bf k}_t$ at the Brewster angle.
It indicates that, in general, the reflection wave and the
refraction wave are not mutually perpendicular. (ii) If
$\mu_x=\mu_z$, this corresponds to the case of isotropic media or
uniaxial materials with the optical axis being $y$-axis. Further,
if $\varepsilon_y=\varepsilon_0$ $({\bf E}_{r0}^e=0)$, then ${\bf
k}_r$ will be perpendicular to ${\bf k}_t$ at the Brewster angle.
Or else, they will be not perpendicular mutually. (iii) If
$\mu_x=\mu_z=\mu_0$ $({\bf E}_{r0}^m=0)$ and
$\varepsilon_y\neq\varepsilon_0$ $({\bf E}_{r0}^e\neq0)$, then
${\bf E}_{r0}\neq0$. That is the reason why TE waves do not
exhibit Brewster angle in reflection on ordinary isotropic
dielectric material.

We next discuss some special cases about Brewster angle for TE
waves. (i) It can be shown that if
$\varepsilon_y/\mu_x=\varepsilon_0/\mu_0\cap\mu_x\mu_z\neq\mu_0^2$
, the Brewster angle is $\theta_B^{TE}=0$. (ii)  If
$\varepsilon_y/\mu_x=\varepsilon_0/\mu_0\cap\mu_x\mu_z=\mu_0^2$ ,
then ${\bf E}_{r0}^{e}+{\bf E}_{r0}^{m}\equiv0$ and an arbitrary
angle of incidence will be the Brewster angle. Consequently, the
omnidirectional total transmission occurs, which may lead to
important applications in optics. (iii) If
$\varepsilon_y/\mu_x\neq\varepsilon_0/\mu_0\cap\mu_x\mu_z=\mu_0^2$
, then ${\bf E}_{r0}^{e}+{\bf E}_{r0}^{m}\neq0$ and the Brewster
angle will not exist.

(2).~TM waves. We can discuss Brewster angle of TM waves and come
to conclusions similar to those about the Brewster angle of TE
waves, simply interchanging $\mu_0$ and $\varepsilon_0$,
$\boldsymbol{\mu}$ and $\boldsymbol{\varepsilon}$, respectively.
In addition, it is clear from Eq.~(\ref{E_r=0_H1}) that if
$\mu_y=\mu_0$ $({\bf H}_{r0}^e=0)$, $\varepsilon_x=\varepsilon_z$
and ${\bf k}_r\bot{\bf k}_t$ $({\bf H}_{r0}^m=0)$, then ${\bf
H}_{r0}=0$. Hence, ${\bf k}_r$ is always perpendicular to ${\bf
k}_t$ at the Brewster angle for an isotropic dielectric material.
Further, we can write the Brewster angle as the well-known form
$\theta_B^{TM}=\tan^{-1}{(n'/n)}$, where
$n=\sqrt{\varepsilon_0\mu_0}$ and $n'=\sqrt{\varepsilon\mu}$ are
the indices of refraction of the vacuum and the isotropic
material, respectively. That is how TM waves exhibit Brewster
angle in reflection on an isotropic nonmagnetic material. And the
explanation at the beginning of the paper is practically that
${\bf H}_{r0}=0$.

In conclusion, the origin of Brewster angle for TE (TM) waves is
that the reflected fields generated by the anisotropic material's
electric and magnetic dipoles disappear in the vacuum, i.e., ${\bf
E}_{r0}^e+{\bf E}_{r0}^m=0$ (${\bf H}_{r0}^e+{\bf H}_{r0}^m=0$).

\section{Brewster angle in internal reflection}
Now, let us consider a different situation: light impinges from
the material into vacuum, where the Brewster angle can also occur
\cite{Zhou2003,Grzegorczyk2005}. One may wonder why the Brewster
angle can exist here since there does not exist any dipoles in the
vacuum. In the following, we discuss the mechanism of Brewster
angle in internal reflection.

\subsection{Internal reflection}
Let us consider a plane wave with ${\bf E}_1={\bf E}_{10}
\exp{(i{\bf k}_{1}\cdot{\bf r})}$ and ${\bf H}_1={\bf H}_{10}
\exp{(i{\bf k}_{1}\cdot{\bf r})}$ incident from an anisotropic
material into the vacuum, where ${\bf k}_{1}=k_{1x} \hat{{\bf
x}}+k_{1z}\hat{{\bf z}}$. The polarization ${\bf P}$ and the
magnetization ${\bf M}$  are related to the incident field as
${\bf P}=\varepsilon_0\boldsymbol{\chi}_e\cdot{\bf E}_1$ and ${\bf
M}=\boldsymbol{\chi}_m\cdot{\bf H}_1$, respectively.

\begin{figure}[t]
\centering
\includegraphics[width=8cm]{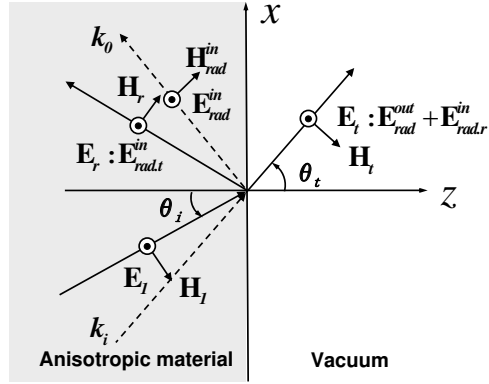}
\caption{\label{shuFig5}Schematic diagram for how the reflected
and transmitted fields are generated for TE waves incident from an
anisotropic material into vacuum. The process is distinct from
those in external reflection as in Fig.~\ref{shuFig2}. Under the
action of the incident field ${\bf E}_1$, the dipoles produce an
internal radiated field ${\bf E}_{rad}^{in}$ in the material and
an external one ${\bf E}_{rad}^{out}$ in the vacuum. ${\bf
E}_{rad}^{in}$ can be regarded as an incident wave from the vacuum
into the material. It produces a reflected wave ${\bf
E}_{rad.r}^{in}$ in the vacuum and a transmitted wave ${\bf
E}_{rad.t}^{in}$ in the material. The magnitude of the real
reflected field is ${\bf E}_{r0}={\bf E}_{rad.t}^{in}=T{\bf
E}_{rad}^{in}$. The final transmitted field is ${\bf E}_{t0}={\bf
E}_{rad.r}^{in}+{\bf E}_{rad}^{out}$. When ${\bf
E}_{rad}^{in}={\bf E}^e_{rad}+{\bf E}^m_{rad}=0$, ${\bf E}_{r0}=0$
and Brewster angle occurs.}
\end{figure}

Following the molecular theory, the incident field ${\bf E}_{1}$
will create a radiated field ${\bf E}^{in}_{rad}$ inside the
material and another radiated field ${\bf E}^{out}_{rad}$ outside
the material. Please see Fig.~\ref{shuFig5}. Now, let us calculate
the radiated fields. First, we need to calculate the Hertz vectors
\begin{equation}\label{Pi_rad_e_internal}
{\bf \Pi}_e=\left\{
\begin{array}{cc}
\displaystyle  \frac{\boldsymbol{\chi}_e\cdot{\bf
E}_{10}\exp{(ik_{1x}x-ik_{0z}z)}}{2k_{0z}(k_{0z}+k_{1z})}+\frac{\boldsymbol{\chi}_e\cdot{\bf
E}_{10}\exp{(i{\bf k}_{1}\cdot{\bf r})}}{k_{t}^2-k_{i}^2},&-\infty<z<0\\
\displaystyle  -\frac{\boldsymbol{\chi}_e\cdot{\bf
E}_{10}\exp{(i{\bf k}_{0}\cdot{\bf
r})}}{2k_{0z}(k_{0z}-k_{1z})},&0\leq z<\infty
\end{array}\right.
\end{equation}
\begin{equation}\label{Pi_rad_m_internal}
{\bf \Pi}_m=\left\{
\begin{array}{cc}
\displaystyle  \frac{\boldsymbol{\chi}_m\cdot{\bf
H}_{t0}\exp{(ik_{1x}x-ik_{0z}z)}}{2k_{0z}(k_{0z}+k_{1z})}+\frac{\boldsymbol{\chi}_m\cdot{\bf
H}_{t0}\exp{(i{\bf k}_{1}\cdot{\bf r})}}{k_{t}^2-k_{i}^2},&-\infty<z<0\\
\displaystyle  -\frac{\boldsymbol{\chi}_m\cdot{\bf
H}_{t0}\exp{(i{\bf k}_{0}\cdot{\bf
r})}}{2k_{0z}(k_{0z}-k_{1z})},&0\leq z<\infty
\end{array}\right.
\end{equation}
where  ${\bf k}_{0}=k_{1x} \hat{{\bf x}}-k_{0z}\hat{{\bf z}}$ and
$k_{0z}^2=k_i^2-k_{1x}^2$. Then, substituting
Eqs.~(\ref{Pi_rad_e_internal}) and (\ref{Pi_rad_m_internal}) into
Eqs.~(\ref{Ee_dipole_field}) and (\ref{Em_dipole_field}) to
calculate  ${\bf E}^{in}_{rad}$, we come to the following
conclusions: The radiation field in the material is
\begin{equation}\label{E_rad_in}
{\bf E}_{rad}^{in}=-\frac{{\bf k}_0\times[{\bf k}_0\times
(\boldsymbol{\chi}_e\cdot{\bf E}_{10})]}{2k_{0z}(k_{0z}+k_{1z})}
-\frac{{\bf
k}_0\times\{\boldsymbol{\chi}_m\cdot[{\mu_0}{\boldsymbol{\mu}}^{-1}\cdot({\bf
k}_1\times {\bf E}_{10})]\}}{2k_{0z}(k_{0z}+k_{1z})}
\end{equation}
with a phase factor $\exp{(i{\bf k}_{0}\cdot{\bf r})}$; Examining
terms with phase $\exp{(i{\bf k}_{1}\cdot{\bf r})}$, we can see
$k_{ix}=k_{0x}=k_{1x}$ and $k_{iz}=k_{0z}$; We also get the
dispersion relations for TE and TM waves similar to
Eqs.~(\ref{D1}) and (\ref{D2}) with $k_{tx}$ and $k_{tz}$ replaced
by $k_{1x}$ and $k_{1z}$, respectively. In the vacuum, the
external radiation field is
\begin{equation}\label{E_rad=}
{\bf E}_{rad}^{out}=\frac{{\bf k}_i\times[{\bf k}_i\times
(\boldsymbol{\chi}_e\cdot{\bf
E}_{10})]}{2k_{0z}(k_{0z}-k_{1z})}+\frac{{\bf
k}_i\times\{\boldsymbol{\chi}_m\cdot[{\mu_0}{\boldsymbol{\mu}}^{-1}\cdot({\bf
k}_1\times {\bf E}_{10})]\}}{2k_{0z}(k_{0z}-k_{1z})}
\end{equation}
with the phase factor $\exp{(i{\bf k}_i\cdot{\bf r})}$.

Evidently, Eq.~(\ref{E_rad_in}) denotes a vacuum plane wave with a
wave vector ${\bf k}_0$ and can be regarded being incident from
the vacuum into the material. Then, the vacuum wave will be
reflected on the interface and transmitted into the material.
Using the conclusions obtained in Sec.~II, we calculate the
reflection wave ${\bf E}_{rad.r}=R {\bf E}_{rad}^{in}$ and the
transmitted wave ${\bf E}_{rad.t}=T {\bf E}_{rad}^{in}$ where $R$
and $T$ are the reflection and transmission coefficients in
external reflections. The transmitted wave ${\bf E}_{rad.t}$ is
the final reflection wave ${\bf E}_{r0}$,
\begin{equation}\label{E_r_in}
{\bf E}_{r0}=T {\bf E}_{rad}^{in}
\end{equation}
and ${\bf k}_t={\bf k}_i$. For TE wave,  using the dispersion
relation of the material the reflection  coefficient $R_E={
E}_{r0}/{ E}_{10}$ is readily evaluated
\begin{equation}\label{RE_in}
R_{E}=\frac{\mu_0k_{1z}-\mu_xk_{tz}}{\mu_0k_{1z}+\mu_xk_{tz}}.
\end{equation}
The reflection field ${\bf E}_{rad.r}$ are superposed by the
radiation field ${\bf E}_{rad}^{out}$ in the vacuum to produce the
real transmitted wave
\begin{equation}
{\bf E}_{t0}=R {\bf E}_{rad}^{in}+{\bf E}_{rad}^{out}.
\end{equation}
Therefore, the transmission  coefficient $T_E={ E}_{t0}/{E}_{10}$
is obtained as
\begin{equation}
T_{E}=\frac{2\mu_0k_{1z}}{\mu_0k_{1z}+\mu_xk_{tz}},
\end{equation}
and ${\bf k}_r=k_{1x} \hat{{\bf x}}-k_{1z}\hat{{\bf z}}$.
Following a similar way, we can obtain the reflection and
transmission coefficients for TM wave
\begin{equation}\label{RH_in}
R_H=\frac{\varepsilon_0k_{1z}-\varepsilon_xk_{tz}}{\varepsilon_0k_{1z}+\varepsilon_xk_{tz}},~~~
T_H=\frac{2\varepsilon_0k_{1z}}{\varepsilon_0k_{1z}+\varepsilon_xk_{tz}}.
\end{equation}

\subsection{Brewster angle}
Next, we explore the mechanism of Brewster angle in internal
reflection.

In order to satisfy $r_E=|{\bf E}_{r0}/{\bf E}_{10}|^2=0$, it
requires ${\bf E}^{in}_{rad}=0$ in Eq.~(\ref{E_r_in}).
Equation~(\ref{E_rad_in}) is similar to Eq.~(\ref{E_r=}), then one
can obtain conclusions similar to those about Brewster angle for
external reflection in the subsection B of Sec.~II after replacing
$k_{iz}$ and $k_{tz}$ with $k_{tz}$ and $k_{1z}$, respectively.
Thus the conditions for Brewster angle in internal reflection are
identical to  those of external reflection. The Brewster angle of
internal reflection can be obtained by Snell's law
$\sin{\theta_B}=k_{t}\sin{\theta_t}/{k_{1}}$, where $\theta_t$ is
equal to the Brewster angle of external reflection. Therefore, we
know that if the condition
\begin{equation}\label{Brewster_E_condition_in}
0<\frac{1-\frac{\mu_0\varepsilon_y}{\mu_x\varepsilon_0}}{1-\frac{\mu_0\mu_0}{\mu_x\mu_z}}<1
\end{equation}
is satisfied, the Brewster angle for TE waves is
\begin{equation}\label{Brewster_E_in}
\theta_{B}^{TE}=\sin ^{-1}
\sqrt{\frac{\mu_0\mu_z(\varepsilon_y\mu_0-\varepsilon_0\mu_x)}
{\varepsilon_0\mu_0(\mu_x^2-\mu_x\mu_z)+\varepsilon_y\mu_z(\mu_0^2-\mu_x^2)}}.
\end{equation}
Similarly, we conclude that under the condition
\begin{equation}\label{Brewster_H_condition_in}
0<\frac{1-\frac{\varepsilon_0\mu_y}{\varepsilon_x\mu_0}}{1-\frac{\varepsilon_0\varepsilon_0}{\varepsilon_x
\varepsilon_z}}<1,
\end{equation}
there exists a Brewster angle for TM waves
\begin{equation}\label{Brewster_H_in}
\theta_{B}^{TM}=\sin ^{-1}
\sqrt{\frac{\varepsilon_0\varepsilon_z(\mu_y\varepsilon_0-\mu_0\varepsilon_x)}{
\varepsilon_0\mu_0(\varepsilon_x^2-\varepsilon_x\varepsilon_z)+\mu_y\varepsilon_z(\varepsilon_0^2-\varepsilon_x^2)}}.
\end{equation}
Through choosing appropriate material parameters, i.e.,
${\boldsymbol \varepsilon}$ and ${\boldsymbol \mu}$, Brewster
angles can happen to both TE and TM waves. Since the requirements
on ${\boldsymbol \varepsilon}$ and ${\boldsymbol \mu}$ for
Brewster angle in internal reflection are the same as those in
external reflection, we can discuss and come to conclusions about
the Brewster angle in internal reflection similar to in external
reflection.

\section{Conclusion}\label{sec4}
In summary, we have used the extinction theorem to generalize the
existence condition of Brewster angle from the isotropic
dielectric material to the anisotropic dielectric-magnetic
material. We investigated the Brewster angle not only in external
reflection, but also in internal reflection. We found the
mechanism for Brewster effect is that the total contributions of
the anisotropic material's electric and magnetic dipoles to the
reflection fields  are zero. Interestingly, the requirements on
the material parameters ${\boldsymbol \varepsilon}$ and
${\boldsymbol \mu}$ for Brewster angle in internal reflection are
the same as those in external reflection, and the corresponding
Brewster angle is just the refracted angle of external reflection
at the incidence of external Brewster angle. This point is
consistent with the reversibility of light ray. We also discussed
in detail the impact of ${\boldsymbol \varepsilon}$ and
${\boldsymbol \mu}$ on the Brewster angle. We found that, through
choosing appropriate ${\boldsymbol \varepsilon}$ and ${\boldsymbol
\mu}$ the anisotropic material can exhibit Brewster angles for TE
waves, or TM waves, or both. Moreover, the Brewster effect can
happen to TE and TM waves simultaneously and the omnidirectional
total transmission will occur, which may lead to important
applications in practice.

Although based on molecular optics theory, these conclusions are
applicable to metamaterials consisting of SRRs and wires. That is
because both the SRR and the wire dimensions are much smaller than
the wavelength of interest \cite{Shelby2001}. Then the unit cells
of SRR and wire can be modelled as the molecules (or atoms) in
ordinary materials. Actually, Belov \textit{et al.} have used the
Ewald-Oseen extinction theorem to investigate the boundary problem
of metamaterials \cite{Belov2006}. We hope that our results will
provide references in manufacturing materials for specific
purposes, such as making polarization devices.

\begin{acknowledgements}
This work was supported in part by the National Natural Science
Foundation of China (No.~10125521, 10535010) and the 973 National
Major State Basic Research and Development of China (G2000077400).
\end{acknowledgements}

\end{document}